\pdfoutput=1
\documentclass[10pt,twocolumn,english,sort,twocolumn]{IEEEtran}
\usepackage{helvet}

\usepackage[T1]{fontenc}
\usepackage[latin9]{inputenc}
\usepackage[letterpaper]{geometry}
\usepackage{color}
\usepackage{babel}
\usepackage{array}
\usepackage{units}
\usepackage{amstext}
\usepackage{amssymb}
\usepackage{graphicx}
\usepackage{setspace}
\usepackage{esint}
\usepackage[unicode=true,pdfusetitle,
 bookmarks=true,bookmarksnumbered=false,bookmarksopen=false,
 breaklinks=false,pdfborder={0 0 0},backref=false,colorlinks=false]
 {hyperref}

\makeatletter

\providecommand{\tabularnewline}{\\}

\usepackage[comma,super]{natbib}
\renewcommand\@biblabel[1]{$^{#1}$}
\usepackage[modulo]{lineno}
\date{}
\usepackage{remreset}

\makeatother

\begin{document}

\title{Near optimal neural network estimator for spectral x-ray photon counting
data with pileup}

\author{Robert E. Alvarez\thanks{ralvarez@aprendtech.com}}
\maketitle
\begin{abstract}
Purpose: A neural network estimator to process x-ray spectral measurements
from photon counting detectors with pileup. The estimator is used
with an expansion of the attenuation coefficient as a linear combination
of functions of energy multiplied by coefficients that depend on the
material composition at points within the object {[}R.E. Alvarez and
A. Macovski, Phys. Med. Biol., 1976, 733-744{]}. The estimator computes
the line integrals of the coefficients from measurements with different
spectra. Neural network estimators are trained with measurements of
a calibration phantom with the clinical x-ray system. One estimator
uses low noise training data and another network is trained with data
computed by adding random noise to the low noise data. The performance
of the estimators is compared to each other and to the Cramer-Rao
lower bound (CRLB). 

Methods: The estimator performance is measured using a Monte Carlo
simulation with an idealized model of a photon counting detector that
includes only pileup and quantum noise. Transmitted x-ray spectra
are computed for a calibration phantom. The transmitted spectra are
used to compute random data for photon counting detectors with pileup.
Detectors with small and large dead times are considered. Neural network
training data with extremely low noise are computed by averaging the
random detected data with pileup for a large numbers of exposures
of the phantom. Each exposure is equivalent to a projection image
or one projection of a computed tomography scan. Training data with
high noise are computed by using data from one exposure. Finally,
training data are computed by adding random data to the low noise
data. The added random data are multivariate normal with zero mean
and covariance equal to the sample covariance of data for an object
with properly chosen attenuation. 

To test the estimators, random data are computed for different thicknesses
of three test objects with different compositions. These are used
as inputs to the neural network estimators. The mean squared errors
(MSE), variance and square of the bias of the neural networks' outputs
with the random object data are each compared to the CRLB. 

Results: The MSE for a network trained with low noise data and added
noise is close to the CRLB for both the low and high pileup cases.
Networks trained with very low noise data have low bias but large
variance for both pileup cases. Networks trained with high noise data
have both large bias and large variance.

Conclusion: With a properly chosen level of added training data noise,
a neural network estimator for photon counting data with pileup can
have variance close to the CRLB with negligible bias. 

\vspace{0.2cm}

\hspace{-0.1cm}Key Words: spectral x-ray, photon counting, pileup,
dual energy, Cramèr-Rao lower bound
\end{abstract}

\section{INTRODUCTION}

The estimator is used with the Alvarez-Macovski method\cite{Alvarez1976}.
In this method, the x-ray attenuation coefficient is approximated
as a linear combination of basis functions of energy multiplied by
coefficients that depend only on the material composition at each
point within the object. The estimator computes the line integrals
of the coefficients from transmitted x-ray measurements with different
energy spectra. Recently, neural networks have been suggested for
this application\cite{LeeNeuralNetEstimator2012,ZimmermanPMB2015neuralEsti,touch_neural_PMB_2016}
by using the universal function approximation theorem\cite{CybenkoNeuralNetworkApproximation}
to invert the nonlinear transformation from the the line integrals
to the spectral measurements. 

If the number of spectral measurements is equal to the number of basis
functions then any estimator that inverts the deterministic transformation
is the maximum likelihood estimator and it gives optimal performance\cite{Alvarez1976,Alvarez2011,AlvarezDESolve32015}.
That is, in the limit of large photon counts it is unbiased and has
covariance equal to the Cramer-Rao lower bound (CRLB)\cite{KayV1Ch7_MLE}.
If the number of spectral measurements is greater than the number
of basis functions then simply inverting the transformation does not
necessarily give optimal results. For this case, the optimal estimator
needs to use the probability distribution of the measurement noise
to provide optimal performance\cite{Kay1993a}. 

Systems with more spectral measurements than the number of basis functions
are becoming increasingly important because of the introduction of
photon counting detectors into medical x-ray imaging systems\cite{taguchi2013vision}.
With these detectors, we can use pulse height analysis (PHA) to categorize
the energy of individual photons into separate energy bins. The counts
in each bin constitute a separate spectral measurement and the number
of bins depends on technical factors of the detector design\cite{Knoll2000}
but is typically larger than the basis set dimension. 

There is no explicit way to incorporate the noise probability distribution
with a neural network but in the past noise has been added to the
training data to improve the generalizability of the network\cite{sietsma1988neural,bishop_neural_add_noise_1995,neural_an_add_noise_1996,zur_neural_add_noise_2009}.
For our application, the noise variance varies exponentially with
the variables being estimated and the off diagonal terms of the covariance
also change with these variables\cite{AlvarezSNRwithPileup2014}.
Therefore, it is not clear whether adding noise to the training data
will result in optimal performance and what is the proper level of
noise to provide this performance. These questions are examined in
this paper. 

The approach used is to train the neural network with measurements
of a calibration phantom in the clinical x-ray system. Low noise training
data are computed by averaging together a large number (1000) of exposures
of the phantom. Another set of training data is computed by adding
zero mean multivariate normal random data with a covariance equal
to the values on a properly chosen interior step of the calibration
phantom to the low noise data. Training data with high noise are computed
by using only one exposure. The output noise of neural network estimators
trained with the three data sets are compared to each other and to
the CRLB. Other factors that affect the estimator output noise include
the number of steps in the calibration phantom as well as the architecture
of the network.

Current state of the art photon counting detectors have defects including
incomplete photon energy measurement due to K radiation and Compton
scattered photon escape, charge sharing and trapping, polarization
and other effects\cite{Overdick2008,taguchi2013vision}. As the detector
state of the art improves, we can expect these defects may be reduced
to negligible levels. However, all photon counting detectors have
a finite response time and, since the inter-arrival times of x-ray
photons on the detector sensor are exponentially distributed\cite{Barrett2003Ch11},
there will always be a non-zero probability for two or more photons
to enter the detector during its response time no matter how small.
In addition, x-ray quantum noise is, of course, universal. In order
to focus on these fundamental issues an idealized model that includes
only quantum noise and pileup is used with a Monte Carlo simulation
to test the neural network estimators' performance. 

Lee et al.\cite{LeeNeuralNetEstimator2012} used a neural network
estimator but did not examine noise in the output. Zimmerman and Schmidt\cite{ZimmermanCompareAtable2iterPMB2014,ZimmermanPMB2015neuralEsti}
studied noise but used noisy training data from a single exposure
of the calibration phantom. Touch et al.\cite{touch_neural_SPIEMedimg_2016,touch_neural_PMB_2016}
used a neural network to correct projection data for defects and deadtime
of photon counting detectors. They then reconstructed data from individual
PHA bins to produce images of the object attenuation at a set of different
x-ray energies instead of the basis set coefficient images produced
by the estimator of this paper.

\section{METHODS}

\subsection{The estimation problem\label{sub:A-space-estimator}}

For biological materials and an externally administered high atomic
number contrast agent we need three or more functions to accurately
approximate the attenuation coefficient\cite{alvarez2013dimensionality},
\begin{equation}
\mu(\mathbf{r},E)=a_{1}(\mathbf{r})f_{1}(E)+a_{2}(\mathbf{r})f_{2}(E)+a_{3}(\mathbf{r})f_{3}(E).\label{eq:3-func-decomp}
\end{equation}

\noindent{}In this equation, $a_{i}(\mathbf{r})$ are the basis set
coefficients, $f_{i}(E)$ are the basis functions and the subscripts
are $i=1\ldots3$. As implied by the notation, the basis set coefficients
$a_{i}(\mathbf{r})$ are functions only of the position $\mathbf{r}$
within the object and the basis functions $f_{i}(E)$ are functions
only of the x-ray energy $E$. Contrast agents with more than one
high atomic number element may require additional functions of energy. 

Neglecting scatter, the expected value of the number of transmitted
photons $n_{k}$ for effective spectrum $S_{k}(E)$ is
\begin{equation}
\lambda_{k}=\left\langle n_{k}\right\rangle =\int S_{k}(E)e^{-\int\mu\left(\mathbf{r},E\right)d\mathbf{r}}dE,\ k=1\ldots n_{spect}\label{eq:Ik-integral}
\end{equation}

\noindent{}where $nspect$ is the number of spectral measurements,
$\left\langle \ \right\rangle $ denotes expected value and the integral
in the exponent is on a line from the x-ray source to the detector.
With pulse height analysis the effective spectrum for an energy bin
measurement can be idealized to be 
\begin{equation}
S_{k}\left(E\right)=S_{incident}\left(E\right)\Pi_{k}\left(E\right)\label{eq:PHA-bin-spect}
\end{equation}

\noindent{}where $S_{incident}\left(E\right)$ is the x-ray spectrum
incident on the detector and $\Pi_{k}\left(E\right)$ is a rectangle
function equal to one inside the energy bin and zero outside. 

Using Eq. \ref{eq:3-func-decomp}, the line integral in Eq. \ref{eq:Ik-integral}
can be expressed as 
\begin{equation}
\int\mu\left(\mathbf{r},E\right)d\mathbf{r}=A_{1}f_{1}(E)+A_{2}f_{2}(E)+A_{3}f_{3}(E).\label{eq:L(E)-A1-f1-A2-f2}
\end{equation}

\noindent{}where $A_{i}=\int a_{i}\left(\mathbf{r}\right)d\mathbf{r},\ i=1\ldots3$.
The $A_{i}$ can be summarized as the components of the A-vector $\mathbf{A}$
and the measurements by a vector $\mathbf{N}$ whose components are
the measurements with the effective spectra, $n_{k},\ k=1\ldots n_{spect}$.
Since the x-ray transmission is exponential in $\mathbf{A}$, we can
approximately linearize the measurements by taking logarithms. The
results is the log measurement vector 
\begin{equation}
\mathbf{L}=-\log(\mathbf{N/\mathbf{N_{0}}}),\label{eq:L-definition}
\end{equation}
where $\mathbf{N_{0}}$ is the expected value of the measurements
with no object in the beam and the division means that corresponding
elements of the vectors are divided. 

Eq. \ref{eq:Ik-integral} defines a relationship between $\mathbf{A}$
and the expected value of the measurement vector, $\mathbf{\left\langle L(A)\right\rangle }$.
In an x-ray system, the objective of the estimator is to invert the
relationship with noisy data and to compute the best estimate of the
A-vector taking into account the probability distribution of the noise.

\subsection{The neural network estimator\label{sub:A-neural-network}}

\begin{figure*}
\centering{}\includegraphics{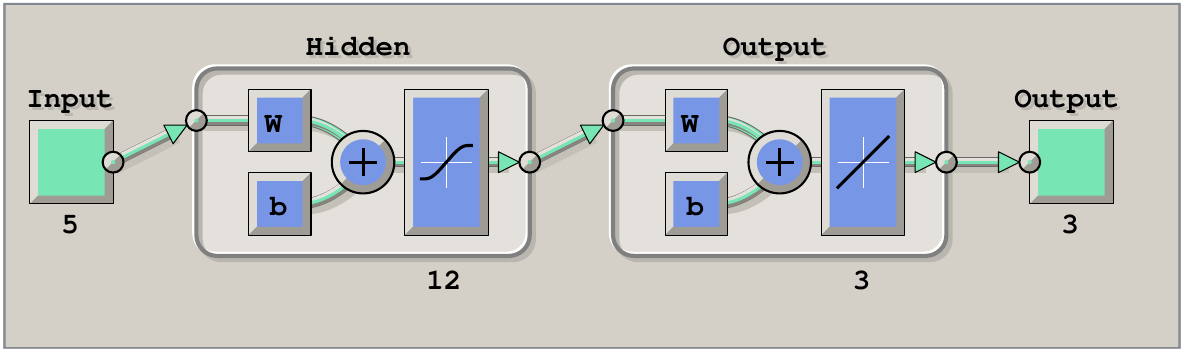}

\protect\caption{Neural network estimator block diagram. One hidden layer with $12$
processing elements and three output elements was found to give near
optimal performance. \label{fig:estimator-block-diagram}}
\end{figure*}

The neural network shown in Fig. \ref{fig:estimator-block-diagram}
was used as an estimator. The inputs to the network are the components
of the $\mathbf{L}$ vector, Eq. \ref{eq:L-definition}. The network
had one hidden layer with $12$ processing elements and three output
elements, which are the estimates of the components of the A-vector,
$\mathbf{A}$. This simple network was found to give near optimal
performance due to the near linearity of the relationship between
$\mathbf{L}$ and $\mathbf{A}.$ Other networks may also give good
performance but they cannot have lower noise than the CRLB.

The hidden elements had a sigmoid response
\[
{\it sigmoid}(x)=\frac{2}{1+e^{-2x}}-1
\]

\noindent{}and the output elements had a linear response. 

The network was trained with calibration data measured with the clinical
x-ray system as discussed in Sec. \ref{sub:A-Calibration-phantom}.
The Levenberg-Marquardt algorithm was used for training. This algorithm
selected the weights and offsets of the network to minimize the mean
squared error of the estimates of the calibration phantom's known
A-vectors using the $\mathbf{L}$ data of the calibrator measured
with the clinical x-ray system. During the training, random subsets
of the data were used for training (70\%), validation (15\%), and
final testing (15\%). The stopping criterion was that the error with
the validation set not decrease for six iterations or the maximum
epochs, 500, were reached. 

The neural network was used with all input data including those that
fall outside the convex hull of the training data.

\subsection{The neural network training data\label{sub:A-Calibration-phantom}}

\begin{figure}
\centering{}\includegraphics[scale=0.5]{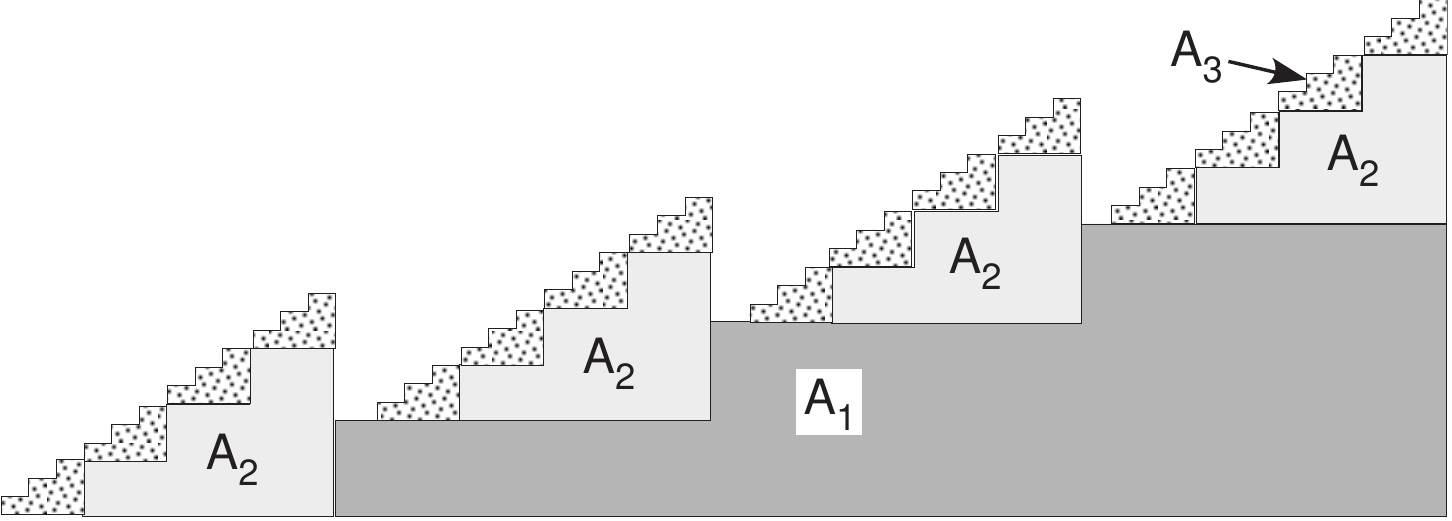}

\protect\caption{Three material calibration phantom. Step wedges of three materials
with known thicknesses have known A-vectors at points in the three
dimensional space. The transmitted flux through the calibration phantom
is measured using the clinical x-ray system. The measurement data
are stored by the system computer in a memory to be used to train
the neural network as discussed in Sec. \ref{sub:A-neural-network}.
See the estimator block diagram in Fig. \ref{fig:estimator-block-diagram}.
\label{fig:3-material-cal-phantom}}
\end{figure}

The neural network was trained using measurements of the calibration
phantom shown in Fig. \ref{fig:3-material-cal-phantom} with the clinical
x-ray system. The figure shows a side-view of a three material phantom.
The purpose of the calibration phantom is to provide values of the
measurement vector \textrm{$\mathbf{L}$} for a set of points in three
dimensional A-space. If we use the attenuation coefficient functions
of the materials of the calibrator as the basis set\cite{Alvarez1979},
then the\textrm{ $\mathbf{A}$ }vectors for each step are simply the
thicknesses of the materials along lines from the x-ray source to
the detector. The phantom can be constructed from stable, machinable
materials such as acrylic plastic and aluminum. The third material
could be, as an example, a plastic resin with molecular linked iodine
whose attenuation simulates iodine contrast agent in blood\cite{qrm-ctiodine}. 

The calibrator data are acquired with the clinical x-ray imaging system
and do not require additional physics instruments that may be unavailable
in clinical institutions. With a fan beam computed tomography system,
the actual path lengths through different parts of phantom can be
computed from its dimensions and the known geometry of the x-ray system
by developing a method to locate the calibrator accurately with respect
the the scanner gantry such as affixing pins to the phantom and scanning
it. Fig. \ref{fig:3-material-cal-phantom} shows a phantom with uniform
steps but exponentially spaced thicknesses were used in the Monte
Carlo simulations. They provide better results since they have closer
spaced samples in the region near the origin where the gradient of
\textrm{$\mathbf{L(A)}$} is highest.

\subsection{Training with noisy data\label{sub:Training-with-noisy-data}}

The Monte Carlo simulation described in Sec. \ref{sub:Monte-Carlo-simulation}
was used to study the optimal training data noise level. The noise
was varied by adjusting the number of exposures averaged together
for the training set where one exposure had the number of photons
used for a clinical object image. Training data with low noise were
computed by averaging 1000 trials. Alternate training data were computed
by adding pseudo-random computer generated noise to the low noise
data. The added data had a multivariate normal distribution with zero
expected value vector. The covariance was the value for the photon
counting detector data an A-vector with components equal to 0.05 of
the maximum calibrator phantom thicknesses for each of the three materials.
This covariance could be measured experimentally from the sample covariance
for data from the calibration phantom for a step with this A-vector.
Finally, high noise training data were computed by using only one
exposure.

\subsection{Model for photon counting data with pulse pileup\label{sub:Models-for-pulse-pileup}}

\textcolor{black}{The idealized model for photon counting data with
pileup is described in detail in a previous paper\cite{AlvarezSNRwithPileup2014}
and is summarized here. The response of photon counting detectors
with pileup to an incident photon is modeled with the dead time parameter,
$\tau$\cite{Knoll2000}, which is the minimum time between two photons
that are recorded as separate events. Two models are commonly used
to describe the recorded counts with pileup: paralyzable and non-paralyzable.
The non-paralyzable model is used in this paper. Measurements by Taguchi
et al.\cite{taguchi_modeling_2011} indicate it is more accurate at
higher count rates with their detectors and it also leads to simpler
analytical results\cite{yu_fessler_PMB_2000}.}

\textcolor{black}{A model is also needed for the recorded energies
with pileup. One approach is to assume that the recorded energy is
proportional to the integral of the sensor charge pulses during the
dead time\cite{taguchi_modeling_2011}. An idealization of this model
was used that assumes that the recorded energy is the sum of the energies
of the photons that arrive during the dead time regardless of how
close the arrival time of a photon to the end of the period\cite{wang_pulse_2011}.
The idealized model assumes that the photon energy is converted completely
into charge carriers so there are no losses due to Compton or Rayleigh
scattering and all K-fluorescence radiation is re-absorbed within
the sensor. All of the carriers are assumed to be collected so that
there is no charge trapping or charge sharing with nearby detectors. }

\subsection{Probability distribution of pulse height analysis data with pileup\label{sub:PHA-stats}}

Using the idealized pulse pileup model described in the previous Sec.
\ref{sub:Models-for-pulse-pileup}, the probability distribution of
PHA data with pileup is modeled as multivariate normal with the expected
value and covariance in Table \ref{tab:PHA-formulas-1}\cite{wang_pulse_2011,AlvarezSNRwithPileup2014}.
The table has two columns, the first for data with no pileup and the
second with pileup. In the table, the subscript 'rec' denotes recorded
data with pileup. The multivariate normal model can be shown to be
accurate for the number of x-ray photons required for the measurements
in material selective imaging\cite{AlvarezSNRwithPileup2014}.

In the table, 
\begin{equation}
\lambda=\int S_{incident\_det}(E)dE\label{eq:Expected-total-incident}
\end{equation}
\noindent{}is the expected value of the number of photons incident
on the detector during the measurement time, $T$, and $\rho=\nicefrac{\lambda}{T}$
is the average rate of photon arrivals. The pileup parameter, $\eta=\rho\tau$,
is the expected number of photons arriving during a dead time period,
$\tau$. The total number of photons recorded by the detector in all
PHA bins is $N_{rec}$ and the number recorded in PHA bin $k$ is
$N_{rec,k}$. Notice that with non-zero pileup parameter the non-Poisson
factor $D$ is not zero so the counts are not Poisson distributed.
The formulas in the table were validated by a Monte Carlo simulation\cite{AlvarezSNRwithPileup2014}. 

\begin{table*}
\protect\caption{PHA probability distribution. $\lambda$ is the expected number of
photons incident on the detector during the measurement time, $\eta$
is the expected number of photons arriving during a dead time period,
$N_{rec}$ is the total number of photons recorded by the detector
in all PHA bins and $N_{rec,k}$ is the number recorded in PHA bin
$k$. \label{tab:PHA-formulas-1}}

\centering{}%
\begin{tabular}{|c|c|c|c|}
\hline 
\multicolumn{1}{|c}{} & no pileup &  & with pileup\tabularnewline
\hline 
\multicolumn{1}{|c}{$\lambda=\int S_{incident\_det}(E)dE$} & $\left\{ 0<E_{1}<\ldots<E_{nbins}\right\} $ &  & \tabularnewline
\hline 
\hline 
photon number spectrum & $S(E)$ &  & $S_{rec}(E)=\lambda p_{rec}(E)$\tabularnewline
\hline 
normalized spectrum & $p(E)=S(E)/\lambda$ &  & $p_{rec}(E)=\sum_{k=0}^{\infty}\frac{\eta^{k}}{k!}e^{-\eta}\left(p^{(k)}*p\right)$\tabularnewline
\hline 
bin probabilities & $P_{k}=\int_{E_{k-1}}^{E_{k}}p(E)dE$ &  & $P_{rec,k}=\int_{E_{k-1}}^{E_{k}}p_{rec}(E)dE$\tabularnewline
\hline 
expected total counts & $\left\langle N\right\rangle =\lambda$ &  & $\left\langle N_{rec}\right\rangle =\frac{\lambda}{1+\eta}$\tabularnewline
\hline 
variance total counts & $var(N)=\lambda$ &  & $var(N_{rec})=\frac{\lambda}{\left(1+\eta\right)^{3}}$\tabularnewline
\hline 
non-Poisson factor & $D=0$ &  & $D=var(N_{rec})-\left\langle N_{rec}\right\rangle $\tabularnewline
\hline 
expected bin counts & $\left\langle N_{k}\right\rangle =\lambda P_{k}$ &  & $\left\langle N_{rec,k}\right\rangle =\left\langle N_{rec}\right\rangle P_{rec,k}$\tabularnewline
\hline 
variance bin counts & $var(N_{k})=\lambda P_{k}$ &  & $var(N_{rec,k})=\left\langle N_{rec}\right\rangle P_{rec,k}+DP_{rec,k}^{2}$\tabularnewline
\hline 
covariance bin counts & $0$ &  & $cov(N_{j},N_{k})_{j\neq k}=P_{rec,j}P_{rec,k}D$\tabularnewline
\hline 
 &  &  & \tabularnewline
\hline 
\end{tabular}
\end{table*}

\textcolor{black}{If the probability of a zero recorded photon count
value is negligible, as is the case with the large expected values
of counts required for material selective imaging, the parameters
of the normal distribution of the logarithm $\mathbf{L}$ data\cite{PapoulisAthanasios1965}
can be computed from the formulas in Table \ref{tab:PHA-formulas-1}
using:
\begin{equation}
\begin{array}{ccc}
\left\langle log(\mathbf{N})\right\rangle  & = & log(\left\langle \mathbf{N}\right\rangle )\\
var\left(log(\mathbf{N})\right) & = & \frac{var\left(\mathbf{N}\right)}{\left\langle \mathbf{N}\right\rangle ^{2}}\\
cov\left(log(\mathbf{N_{1}}),log(\mathbf{N_{2}})\right) & = & \frac{cov(\mathbf{N_{1},N_{2}})}{\left\langle \mathbf{N_{1}}\right\rangle \left\langle \mathbf{N_{2}}\right\rangle }.
\end{array}\label{eq:log-formuals}
\end{equation}
 }

\subsection{The CRLB for A-vector noise with pileup\label{sub:CRLB-with-pileup}}

The CRLB is the minimum covariance for any unbiased estimator and
is a fundamental limit from statistical estimator theory. It is the
inverse of the Fisher information matrix $\mathbf{F}$ whose elements
are\cite{KayV1Chapter3} 
\begin{equation}
F_{ij}=-\left\langle \frac{\partial^{2}\mathcal{L}}{\partial A_{i}\partial A_{j}}\right\rangle \label{eq:F-element}
\end{equation}

\noindent{}where $\mathcal{L}$ is the logarithm of the likelihood
and the symbol $\left\langle \ \right\rangle $ denotes the expected
value. Kay\cite{KayV1Sec4_5} shows that the Fisher information matrix
for multivariate normal data with expected value $\mathbf{\left\langle L(A)\right\rangle }$
and covariance $\mathbf{C_{L}}$ has elements
\begin{equation}
\begin{array}{ccc}
F_{ij} & = & \left[\frac{\partial\mathbf{\left\langle L(A)\right\rangle }}{\partial A_{i}}\right]^{T}\mathbf{C_{L}^{-1}}\left[\frac{\partial\mathbf{\left\langle L(A)\right\rangle }}{\partial A_{j}}\right]+\ldots\\
 &  & \frac{1}{2}tr\left[\mathbf{C_{L}^{-1}}\frac{\partial\mathbf{C_{L}}}{\partial A_{i}}\mathbf{C_{L}^{-1}}\frac{\partial\mathbf{C_{L}}}{\partial A_{j}}\right]
\end{array}\label{eq:CRLB-gen}
\end{equation}

\noindent{}where the $tr\left[\right]$ is the trace of a matrix.

The CRLB was computed numerically by approximating the derivatives
in Eq. \ref{eq:CRLB-gen} from the first central difference. For example,
to compute $\mathbf{\Delta L}$ we first compute the spectra through
the object with attenuation $\mathbf{A}$ and then with $\mathbf{A+}\Delta\mathbf{A}$.
The transmitted spectra are not affected by pileup since they occur
before the measurement. These transmitted spectra are then used to
compute the expected values of the measurements with pileup using
the formulas in Sec. \ref{sub:PHA-stats}. We can similarly compute
the difference of the covariance of the log data $\mathbf{C_{L}}$using
the formulas in that section.

\subsection{The test object for Monte Carlo simulation\label{sub:test-object}}

\begin{figure}
\centering{}\includegraphics[scale=0.6]{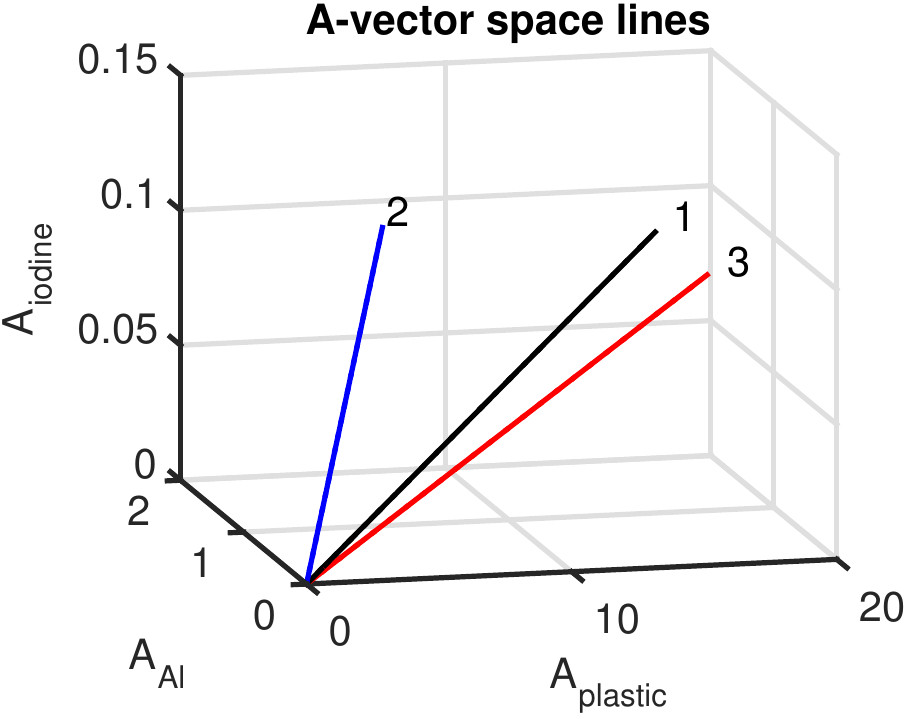}

\protect\caption{Three lines in A-vector space used in the Monte Carlo simulation.
Each line is the A-vectors of different thicknesses of a material
with a specific $\mathbf{a}$ vector of coefficients in its attenuation
coefficient expansion. \label{fig:Three-lines}}
\end{figure}

The performance of the neural network estimator was tested with objects
with A-vectors on three lines through the A-vector space shown in
Fig. \ref{fig:Three-lines}. These correspond to a set of thicknesses
of three uniform objects with different compositions. Each material
has different basis set coefficients in its attenuation coefficient
expansion, Eq. \ref{eq:3-func-decomp}. Summarizing the coefficients
as the components of the $\mathbf{a}$ vector, the line integrals
are therefore $\mathbf{A=a}W$, where $W$ is the object thickness
with units corresponding to attenuation coefficient, for example $g/cm^{2}$.
The A-vectors for different thicknesses of a material therefore fall
on a straight line through the origin in A-vector space. The end points
of the lines used in the simulations, which also specify the ratios
of the $\mathbf{a}$ vector coefficients, were $\left[16,\ 1.2,\ 0.1\right]$,
$\left[5,\ 0.9,\ 0.1125\right]$, and $\left[16,\ 0.375,\ 0.1\right]$
$g/cm^{2}$. Fig. \ref{fig:Three-lines} shows a three dimension plot
of the lines.

\subsection{Random data for Monte Carlo simulation\label{sub:Monte-Carlo-simulation}}

The Monte Carlo simulation compared the mean squared error, the variance,
and the square of the bias of the A-vector estimates to the Cramèr-Rao
lower bound. The generation of the random data for the simulation
started with the computation of a 120 kilovolt x-ray tube spectrum
using the TASMIP algorithm of Boone and Seibert\cite{Boone1997}.
The number of photons incident on the object for each detector element
or pixel was set to $10^{6}$. The measurement time was assumed to
be 10 milliseconds so the rate of photons incident on the detector
with no object in the beam was $10^{8}$ photons per second. Two cases
of pileup were computed by setting the detector dead times to 10 and
1 nanoseconds. These result in zero object thickness pileup parameters,
$\eta_{0}$, of 1 and 0.1 respectively. 

As discussed in Sec. \ref{sub:A-Calibration-phantom}, a basis set
consisting of the attenuation coefficients of acrylic plastic, aluminum,
and an iodine contrast agent simulant consisting of 20\% fraction
by weight iodine in paraffin ($C_{31}H_{64}$ molecular composition)
was used as the basis functions of energy in Eq. \ref{eq:3-func-decomp}.
With this choice, the A-vectors were the thicknesses of each of the
materials in the calibration phantom\cite{Alvarez1979}. The attenuation
coefficients of the materials were computed as the fraction by weight
of each element in its chemical formula multiplied by the attenuation
coefficient of that element. The elements' attenuation coefficients
were computed by piece-wise continuous Hermite polynomial interpolation
of the standard Hubbell-Seltzer tables\cite{Hubbell1999}.

A calibration phantom with 30 steps for each material was implemented.
The thicknesses were geometrically spaced from zero to 20, 1.5 and
0.125 $g/cm^{2}$ for each of the calibration materials respectively.
These were chosen to be greater than the object values so all $\mathbf{L}$
measurements except for noise will fall within the calibration data
convex hull.

For a single A-vector on one of the lines in Fig. \ref{fig:Three-lines},
the TASMIP x-ray tube spectrum was used with Eq. \ref{eq:Expected-total-incident}
to compute the spectrum and the expected value of the total number
of transmitted photons incident on the detector sensor during the
measurement time. The sensor was assumed to be perfectly absorbing
so the signal for each photon was proportional to the photon energy.
The recorded energy with pulse pileup was computed using the pileup
model described in Sec. \ref{sub:Models-for-pulse-pileup}. Five bin
PHA was done on the recorded energies. The PHA energy response functions
were computed with an algorithm that maximized the SNR with no pileup
and were assumed to be perfect rectangles. The algorithm for the optimal
bins is described in a previous paper\cite{AlvarezSNRwithPileup2014}. 

The expected values and covariance of the recorded PHA bin counts
were computed from the formulas in Table \ref{tab:PHA-formulas-1}.
These were used to compute the parameters of the multivariate normal
distributed log data $\mathbf{L}$ using Eq. \ref{eq:log-formuals}.
These data were used to test the neural network estimators.

\subsection{Estimator performance}

The estimator performance was computed for the three training data
sets and for the two cases of the dead time parameter, 1 and 10 nanoseconds.
The neural networks were trained as described in Sec. \ref{sub:Training-with-noisy-data}.
The input data to the estimators were computed as described in Sec.
\ref{sub:Monte-Carlo-simulation}. The estimates of the networks were
computed using the same random data as inputs so their output noise
could be directly compared. The mean square error MSE of the estimates
for 2000 trials was computed as 
\begin{equation}
\mathbf{MSE}=\frac{1}{N_{trials}}\sum_{i=1}^{N_{trials}}\left(\hat{\mathbf{A}}-\mathbf{A_{actual}}\right)^{2}\label{eq:MSE-define}
\end{equation}

\noindent{}where $N_{trails}$ is the number of trials, $\mathbf{\text{\ensuremath{\hat{A}}}}$
is the estimate and $\mathbf{\mathbf{A_{actual}}}$ is the actual
A-vector value. Notice that the MSE is a vector quantity with a value
for each component of $\mathbf{A}$. The sample variance and the square
of the sample bias of the estimates were also computed. These were
plotted after normalizing by dividing by the CRLB variance.

\section{RESULTS \label{sec:RESULTS}}

\subsection{Mean squared error \label{sub:Mean-Squared-Error-low-pileup}}

\begin{figure*}[!t]
\centering{}(a)\includegraphics[scale=0.55]{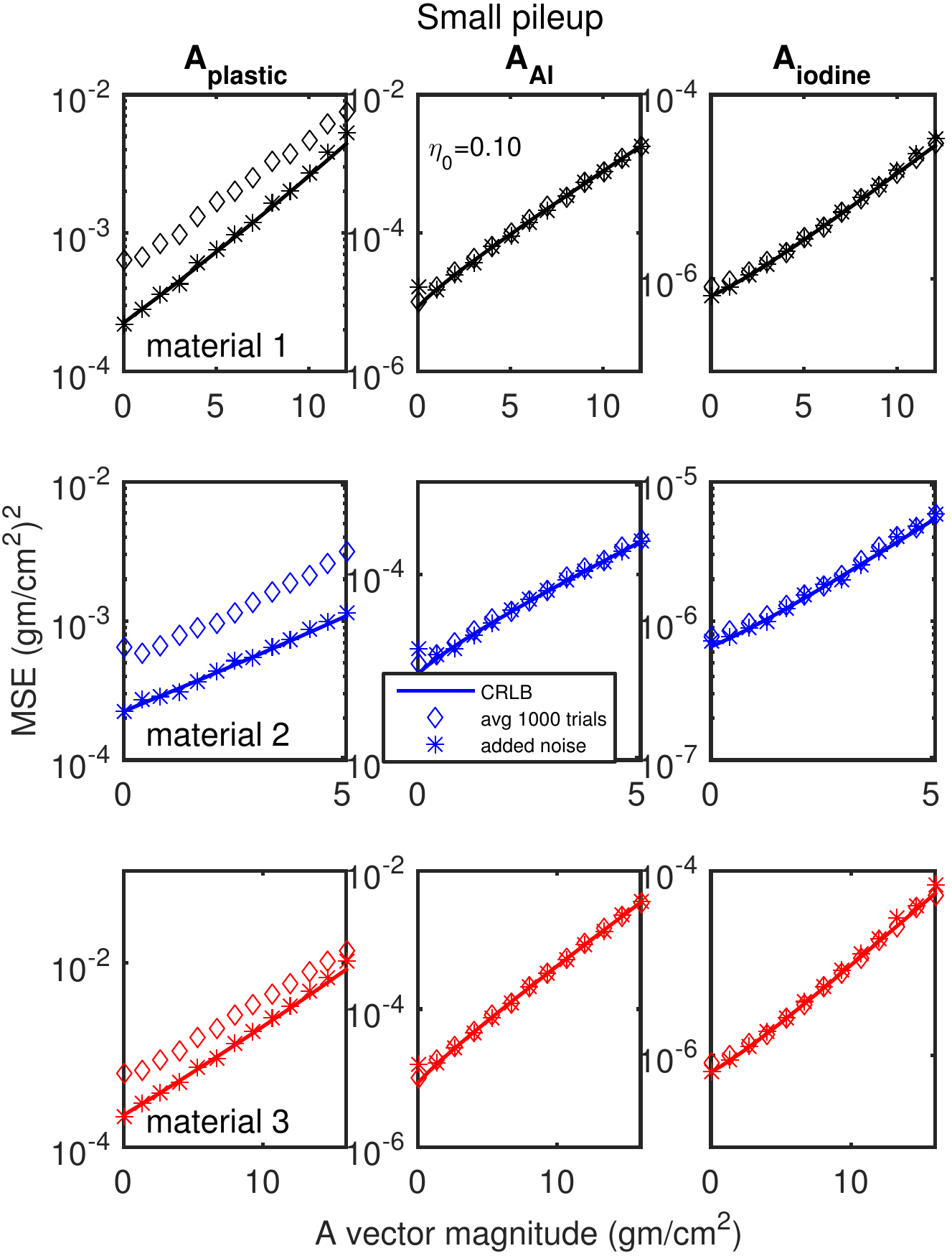}(b)\includegraphics[scale=0.55]{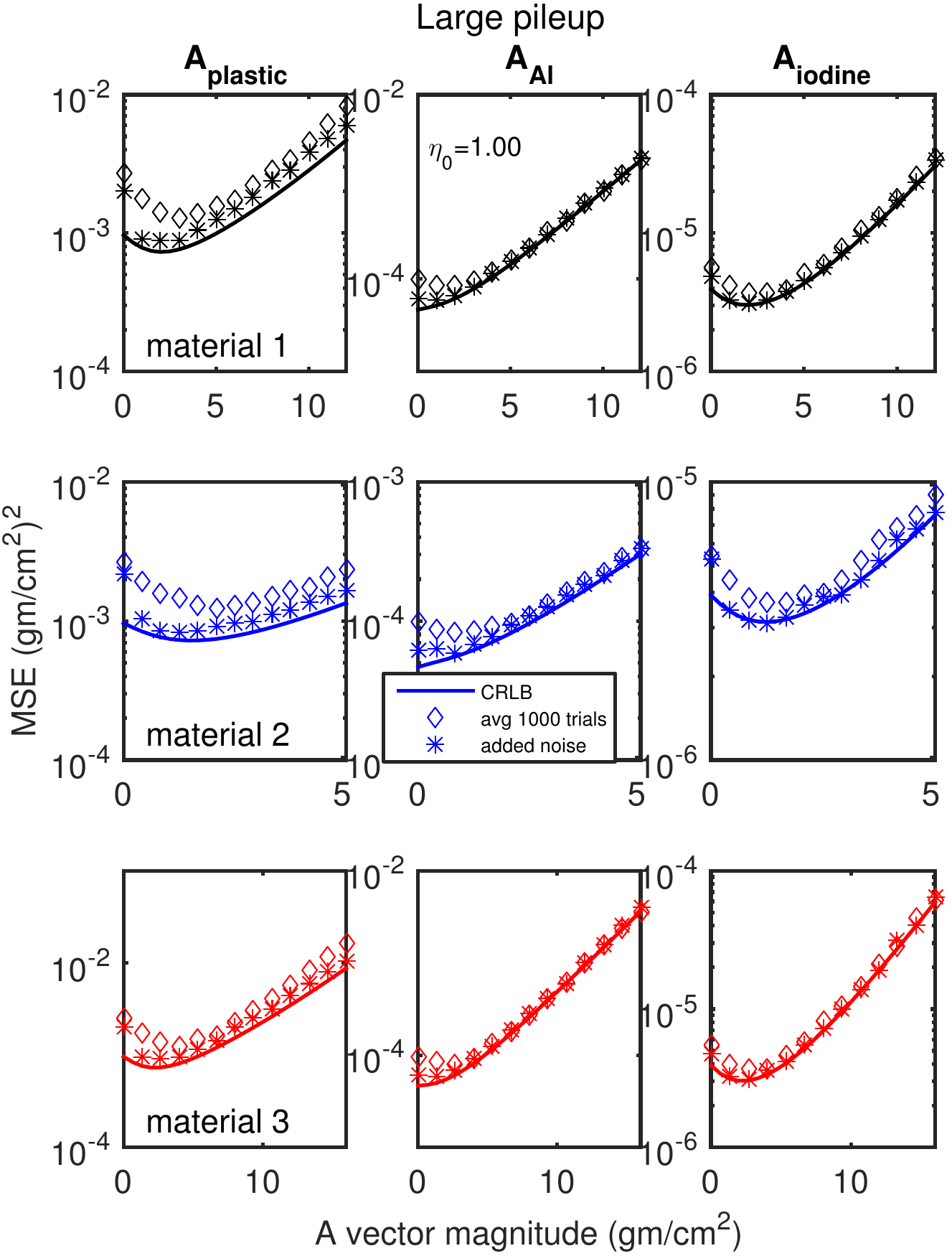}

\protect\caption{Mean squared error vs. A-vector magnitude for the (a) low zero object
thickness pileup case, $\eta_{0}=0.1$, and (b) high pileup case,
$\eta_{0}=1$. The errors are plotted as a function of the A-vector
magnitude for the three lines in Fig. \ref{fig:Three-lines}. The
A-vector magnitude is proportional to object thickness as explained
in Sec. \ref{sub:test-object}. The results for each of the lines
are in separate rows with the A-vector components in separate columns.
The two sets of training data noise were plotted using different symbols:
1000 trials averaged,$\Diamond$ (low noise), and added random noise
$\ast$. \label{fig:MSE}}
\end{figure*}

Fig. \ref{fig:MSE} shows the mean squared errors of the neural network
estimators as a function of A-vector magnitude for the three lines
in Fig. \ref{fig:Three-lines}. Panel (a) has the results for the
low pileup case with zero object thickness dead time parameter, $\eta_{0}=0.1$,
and panel (b) for the high pileup case, $\eta_{0}=1$. The A-vector
magnitude is proportional to object thickness as explained in Sec.
\ref{sub:test-object}. The MSE with the two sets of training data
noise were plotted using different symbols: 1000 trials averaged $\Diamond$
(low noise), added random noise $\ast$. The CRLB variance is plotted
as the solid curves. 

\subsection{Variance \label{sub:Variance}}

\begin{figure*}[p]
\centering{}(a)\includegraphics[scale=0.5]{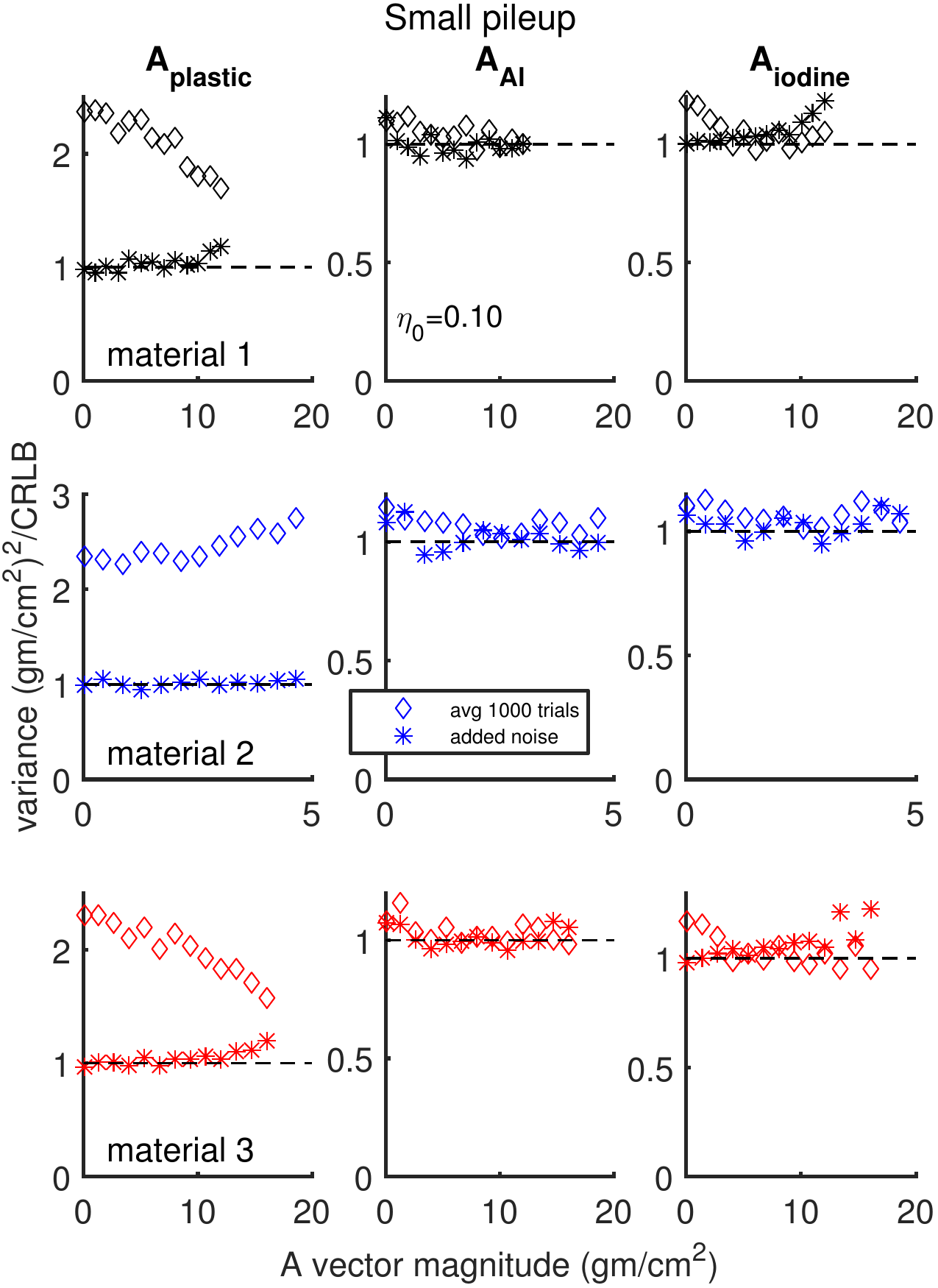}$\ $ (b)\includegraphics[scale=0.5]{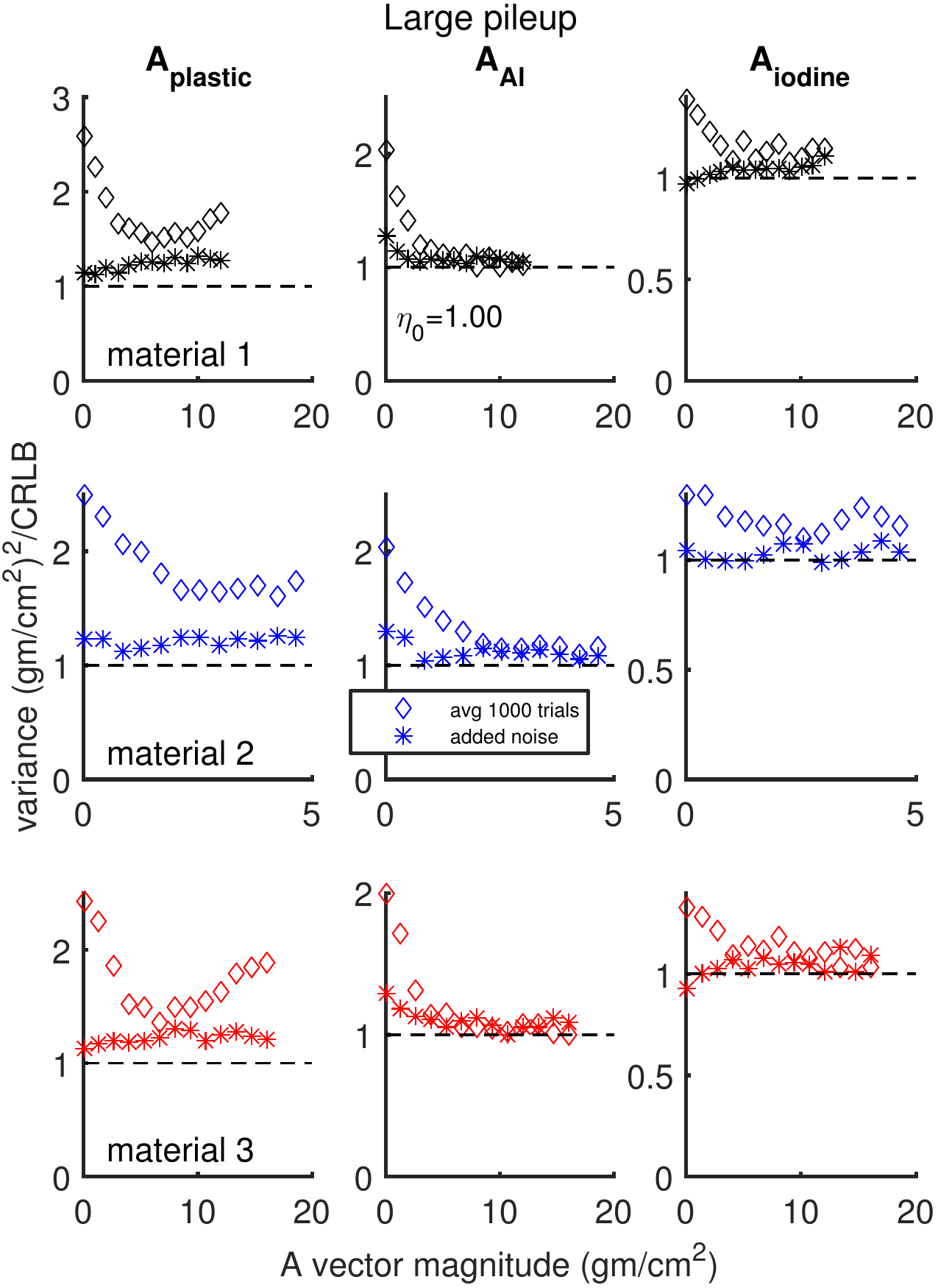}

\protect\caption{Variance for the (a) low zero object thickness pileup case, $\eta_{0}=0.1$,
and (b) high pileup case, $\eta_{0}=1$, plotted as a function of
A-vector magnitude. The symbols are the same as in Fig. \ref{fig:MSE}.
The data are normalized by dividing by the CRLB variance and a dashed
line is drawn at a ratio of one. Notice that the neural network trained
with low noise data has variance substantially larger than the networks
trained with added noise. \label{fig:Var}}
\end{figure*}

The variance of the neural network estimator outputs is shown in Fig.
\ref{fig:Var}. Panel (a) shows the low pileup case, $\eta_{0}=0.1$,
and panel (b) the high pileup case, $\eta_{0}=1$. The symbols for
different training data sets are the same as in Fig. \ref{fig:MSE}.
For each A-vector component, the data are normalized by dividing by
the CRLB variance and a dashed line is drawn at a ratio of one. 

\subsection{Bias}

\begin{figure*}
\centering{}(a)\includegraphics[scale=0.53]{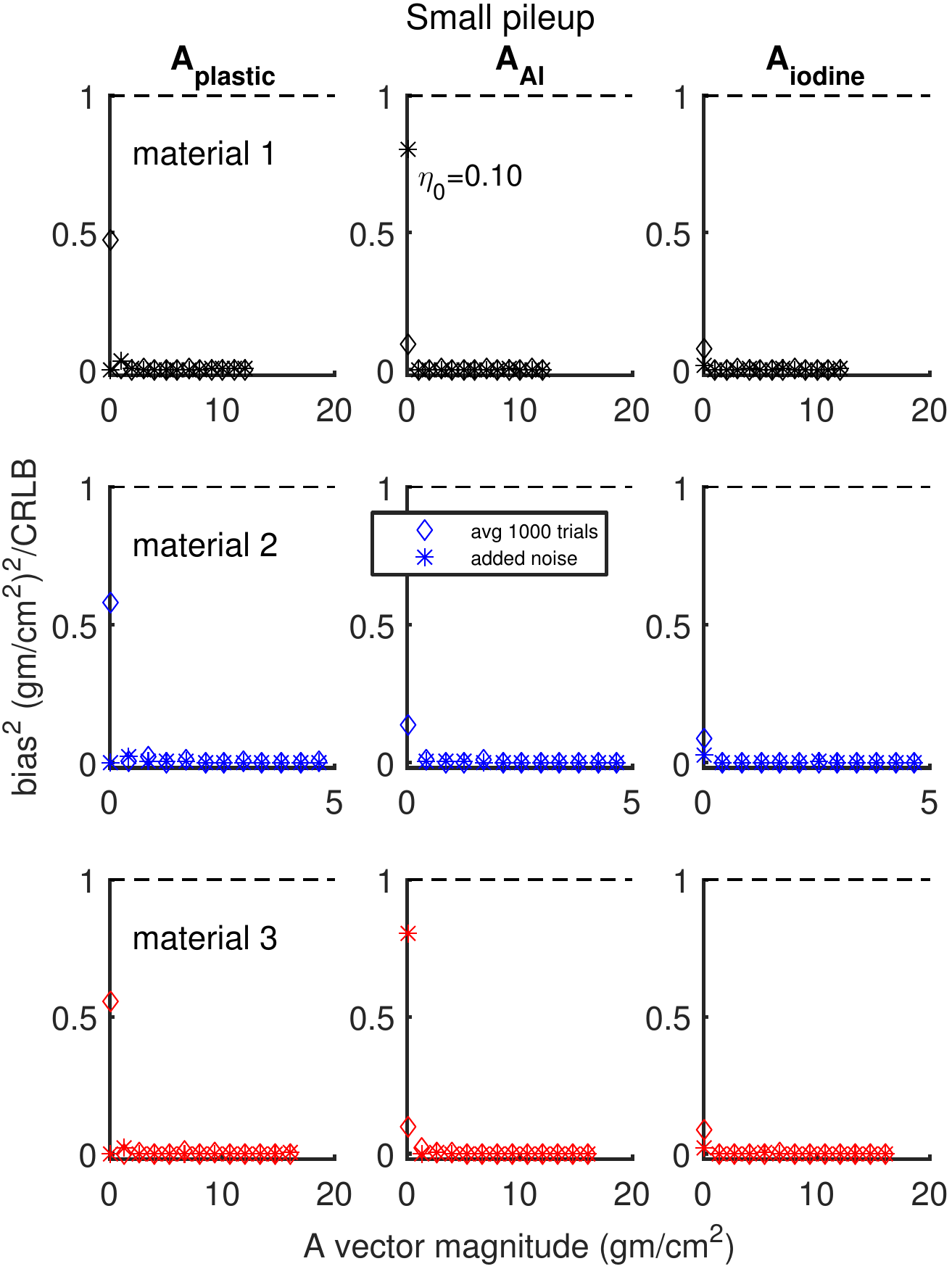}(b) \includegraphics[scale=0.53]{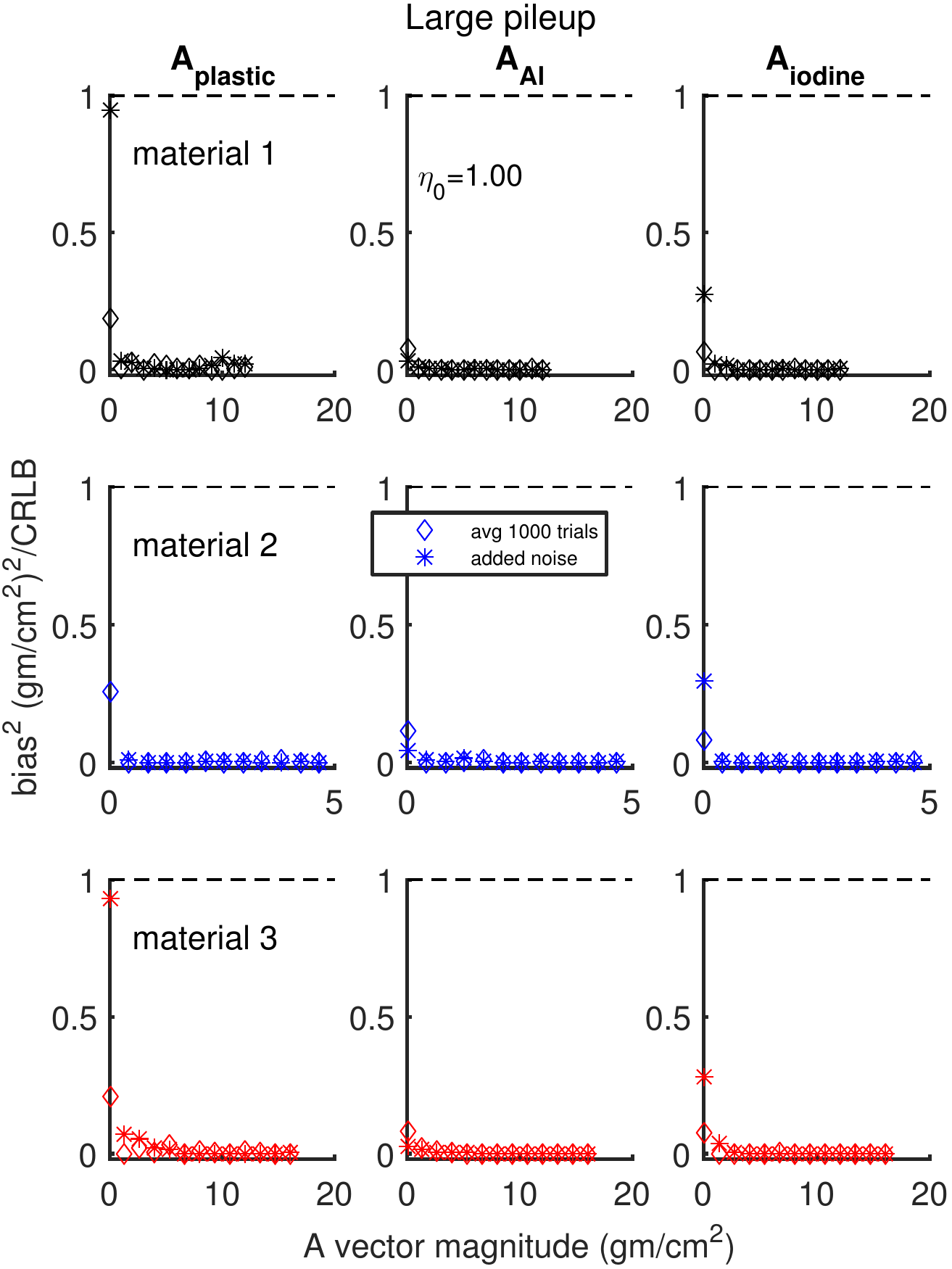}

\protect\caption{The square of the bias for the (a) low pileup case, $\eta_{0}=0.1$
and (b) high pileup case, $\eta_{0}=1$. See the caption of Fig. \ref{fig:Var}
for an explanation of the symbols. The data are normalized by dividing
by the CRLB variance and a dashed line is drawn at a ratio of one.
\label{fig:Bias}}
\end{figure*}

Fig. \ref{fig:Bias} shows the square of the bias for the two pileup
cases. The bias squared data are normalized by dividing by the CRLB
variance and a dashed line is drawn at a ratio of one.

\subsection{Bias and variance of network trained with high noise}

\begin{figure*}
\centering{}(a)\includegraphics[scale=0.5]{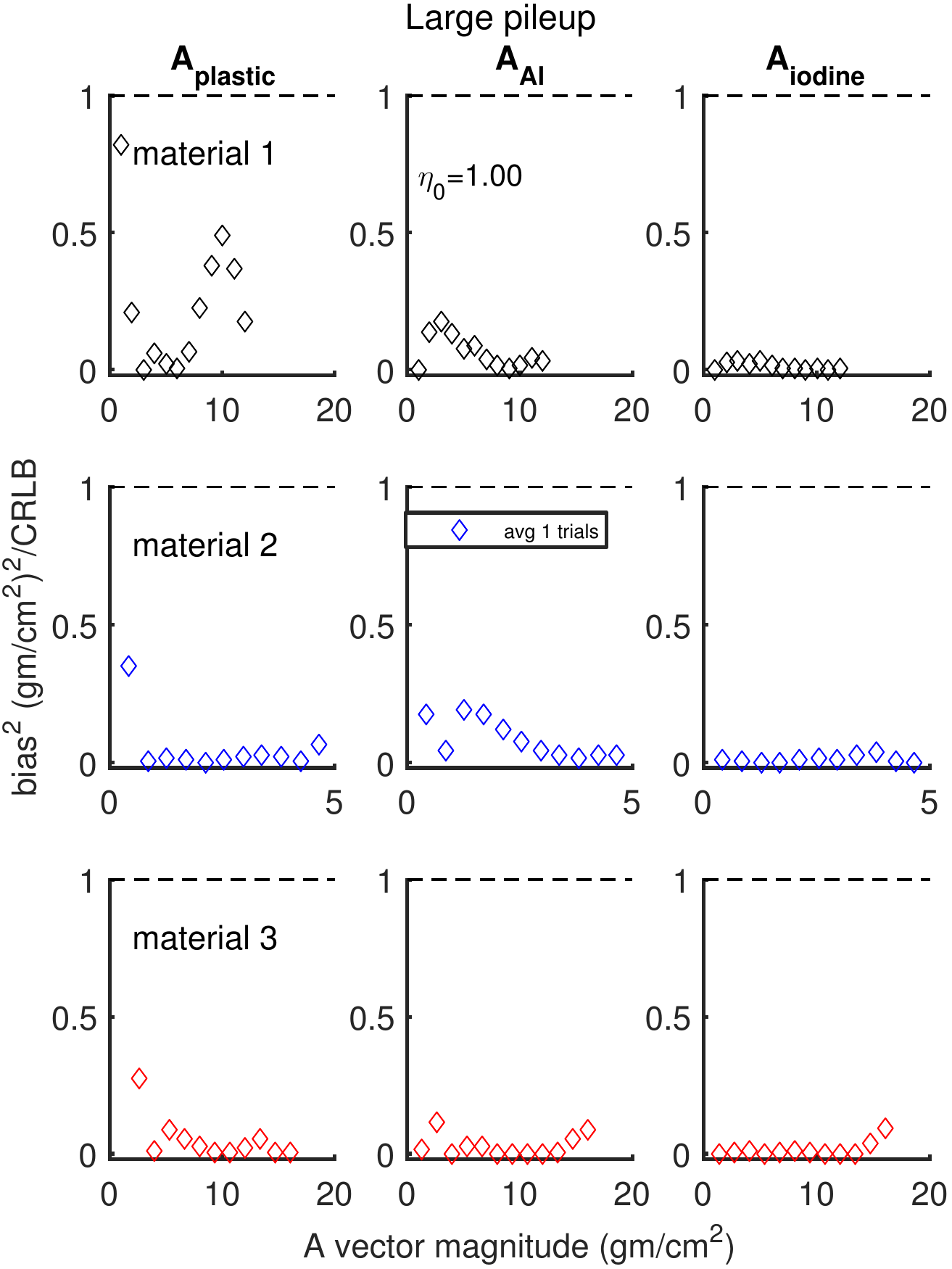}(b)\includegraphics[scale=0.5]{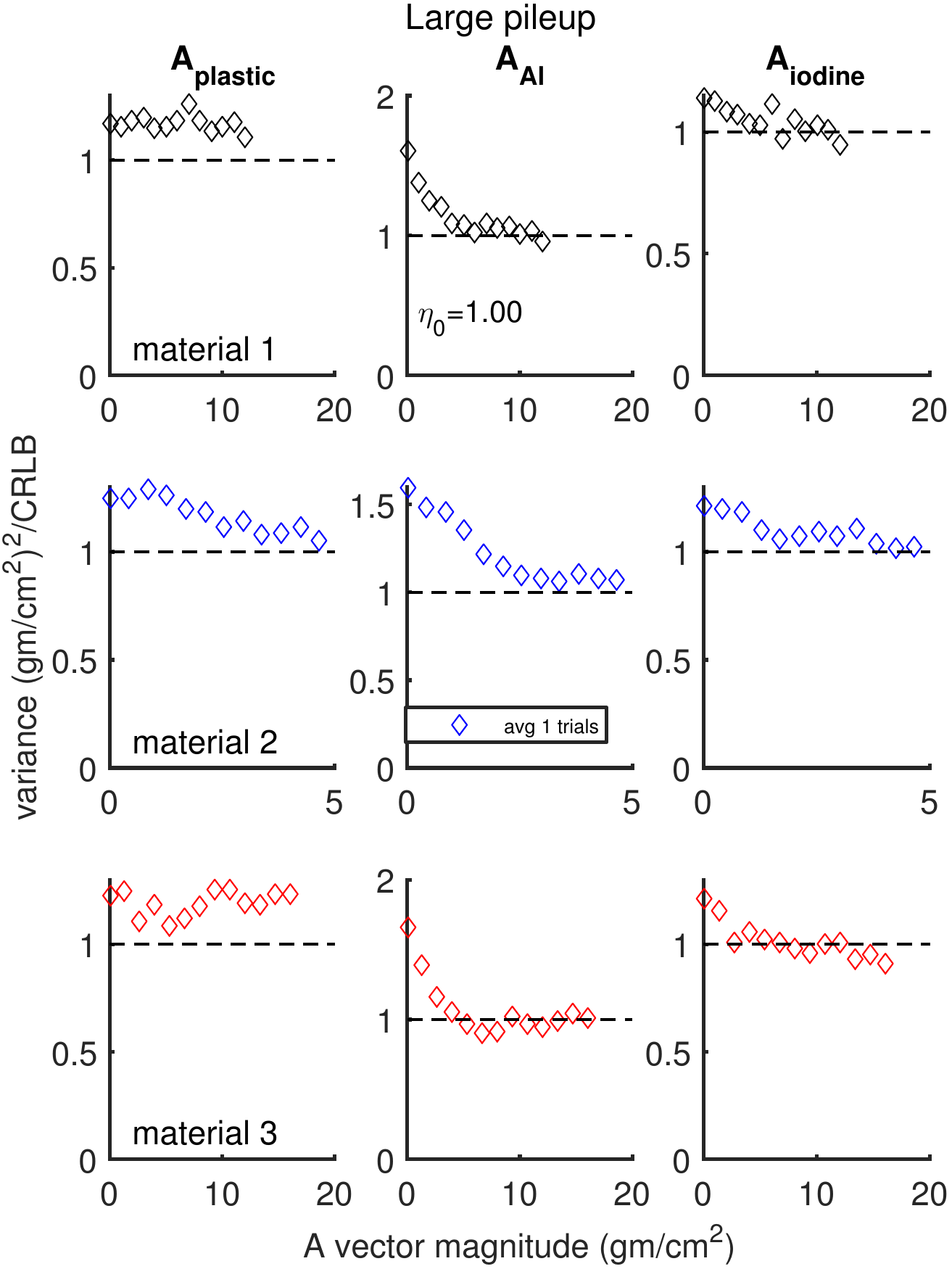}

\protect\caption{Bias (a) and variance (b) of network trained with high noise data
with large pileup parameter. The data are normalized by dividing by
the CRLB variance and a dashed line is drawn at a ratio of one. \label{fig:Bias-var-1-exposure}}
\end{figure*}

Fig. \ref{fig:Bias-var-1-exposure} shows the bias and variance of
a neural network trained with high noise data from a single exposure.
The large pileup case is shown. 

\section{DISCUSSION}

Fig. \ref{fig:MSE} shows that, for either the small or large pileup
cases, the neural network estimator trained with low noise data resulting
from averaging 1000 exposures of the calibration phantom with added
random noise has mean squared errors close to the CRLB. The network
trained with low noise data only has MSE larger than the CRLB.

The mean squared error is equal to the variance plus the square of
the bias and examining each of these components gives us insight into
the effect of training data noise on the performance of the estimators.
The variance results in Fig. \ref{fig:Var} show that the low training
data noise case, plotted as the diamonds has higher variance than
the added noise case. The results in Fig. \ref{fig:Bias} show that
both estimators have low bias. 

Panel (b) of Fig. \ref{fig:MSE} shows that, with large pileup and
for small thicknesses, the CRLB actually decreases as the object thickness
increases. In the absence of pileup, we would expect that the A-vector
noise variance and therefore the CRLB increases exponentially as the
object thickness increases because the number of transmitted photons
incident on the detector during the measurement time decreases exponentially.
With pileup, the arrival rate of photons on the detector and therefore
the pileup parameter also decrease exponentially with object thickness
and, all other factors being equal, a decrease in the pileup parameter
results in lower A-vector noise. The results indicate that in the
small thickness region the decreased noise due to the improvement
in the pileup parameter offsets the increase in noise due to fewer
photons resulting in the decreased MSE shown in the figure. 

Fig. \ref{fig:Bias-var-1-exposure} shows that a neural network trained
with high noise data has both large bias and large variance. 

The small dead time parameters used in the simulation, 10 and 1 nanosecond,
are required to give the desired zero object thickness pileup parameters
due to the large assumed arrival rate of $10^{8}$ photons/second.
This rate, however, may be encountered at the edges of the body in
computed tomography scanners\cite{taguchi2013vision}. The small dead
times may be achievable in the future as sensor materials and detector
electronics improve. Indeed, the performance of some current experimental
systems is approaching these values. For example, Liu\cite{liu2016char-si-det-photon-count2016}
measured approximately 20 nanoseconds dead time for an experimental
silicon strip sensor photon counting detector.

\section{CONCLUSION}

A neural network estimator trained with data with a properly chosen
level of added noise can achieve near CRLB covariance with negligible
bias. Training with low noise data results in low bias but large noise
variance. A network trained with high noise data has both large bias
and large variance. 



\begin{thebibliography}{10}
\providecommand{\url}[1]{#1}
\csname url@samestyle\endcsname
\providecommand{\newblock}{\relax}
\providecommand{\bibinfo}[2]{#2}
\providecommand{\BIBentrySTDinterwordspacing}{\spaceskip=0pt\relax}
\providecommand{\BIBentryALTinterwordstretchfactor}{4}
\providecommand{\BIBentryALTinterwordspacing}{\spaceskip=\fontdimen2\font plus
\BIBentryALTinterwordstretchfactor\fontdimen3\font minus
  \fontdimen4\font\relax}
\providecommand{\BIBforeignlanguage}[2]{{%
\expandafter\ifx\csname l@#1\endcsname\relax
\typeout{** WARNING: IEEEtran.bst: No hyphenation pattern has been}%
\typeout{** loaded for the language `#1'. Using the pattern for}%
\typeout{** the default language instead.}%
\else
\language=\csname l@#1\endcsname
\fi
#2}}
\providecommand{\BIBdecl}{\relax}
\BIBdecl

\bibitem{Alvarez1976}
\BIBentryALTinterwordspacing
R.~E. Alvarez and A.~Macovski, ``{Energy-selective reconstructions in X-ray
  computerized tomography},'' \emph{Phys. Med. Biol.} {\bf 21},  733--44,
  (1976). [Online]. Available:
  \url{http://dx.doi.org/10.1088/0031-9155/21/5/002}
\BIBentrySTDinterwordspacing

\bibitem{LeeNeuralNetEstimator2012}
\BIBentryALTinterwordspacing
W.-J. Lee, D.-S. Kim, S.-W. Kang, and W.-J. Yi, ``Material depth reconstruction
  method of multi-energy x-ray images using neural network,'' \emph{Proc. IEEE
  Engr. Med. Biol. Conf.},''  {\bf } 1514--1517, (Aug 2012). [Online].
  Available: \url{http://dx.doi.org/10.1109/EMBC.2012.6346229}
\BIBentrySTDinterwordspacing

\bibitem{ZimmermanPMB2015neuralEsti}
\BIBentryALTinterwordspacing
K.~C. Zimmerman and T.~G. Schmidt, ``Experimental comparison of empirical
  material decomposition methods for spectral ct,'' \emph{Phys. Med. Biol.}
  {\bf 60}, no.~8,  3175, (2015). [Online]. Available:
  \url{http://dx.doi.org/10.1088/0031-9155/60/8/3175}
\BIBentrySTDinterwordspacing

\bibitem{touch_neural_PMB_2016}
\BIBentryALTinterwordspacing
M.~Touch, D.~P. Clark, W.~Barber, and C.~T. Badea, ``A neural network-based
  method for spectral distortion correction in photon counting x-ray ct,''
  \emph{Phys. Med.Biol.} {\bf 61}, no.~16,  6132, (2016). [Online]. Available:
  \url{http://dx.doi.org/10.1088/0031-9155/61/16/6132}
\BIBentrySTDinterwordspacing

\bibitem{CybenkoNeuralNetworkApproximation}
\BIBentryALTinterwordspacing
G.~Cybenko, ``\BIBforeignlanguage{English}{Approximation by superpositions of a
  sigmoidal function},'' \emph{\BIBforeignlanguage{English}{Mathematics of
  Control, Signals and Systems}} {\bf 2}, no.~4,  303--314, (1989). [Online].
  Available: \url{http://dx.doi.org/10.1007/BF02551274}
\BIBentrySTDinterwordspacing

\bibitem{Alvarez2011}
\BIBentryALTinterwordspacing
R.~E. Alvarez, ``{Estimator for photon counting energy selective x-ray imaging
  with multi-bin pulse height analysis},'' \emph{Med. Phys.} {\bf 38},
  2324--2334, (2011). [Online]. Available:
  \url{http://dx.doi.org/10.1118/1.3570658}
\BIBentrySTDinterwordspacing

\bibitem{AlvarezDESolve32015}
\BIBentryALTinterwordspacing
R.~E. Alvarez, ``{Efficient, non-iterative estimator for imaging contrast
  agents with spectral x-ray detectors},'' \emph{{IEEE} Trans. Med. Imag.},''
  {\bf }(2015). [Online]. Available:
  \url{http://dx.doi.org/10.1109/TMI.2015.2510869}
\BIBentrySTDinterwordspacing

\bibitem{KayV1Ch7_MLE}
S.~M. Kay, \emph{{Fundamentals of Statistical Signal Processing, Volume I:
  Estimation Theory }},\hskip 1em plus 0.5em minus 0.4em\relax Upper Saddle
  River, NJ: Prentice Hall PTR, 1993, Ch. 7.

\bibitem{Kay1993a}
S.~M. Kay, \emph{{Fundamentals of Statistical Signal Processing, Volume I:
  Estimation Theory }},\hskip 1em plus 0.5em minus 0.4em\relax Upper Saddle
  River, NJ: Prentice Hall PTR, 1993.

\bibitem{taguchi2013vision}
\BIBentryALTinterwordspacing
K.~Taguchi and J.~S. Iwanczyk, ``Vision 20/20: Single photon counting x-ray
  detectors in medical imaging,'' \emph{Med. Phys.} {\bf 40},  100901, (2013).
  [Online]. Available: \url{http://dx.doi.org/10.1118/1.4820371}
\BIBentrySTDinterwordspacing

\bibitem{Knoll2000}
G.~F. Knoll, \emph{{Radiation Detection and Measurement}}, 3rd~ed.,\hskip 1em
  plus 0.5em minus 0.4em\relax Hoboken, N.J.: Wiley, 2000.

\bibitem{sietsma1988neural}
\BIBentryALTinterwordspacing
J.~Sietsma and R.~J. Dow, ``Neural net pruning-why and how,'' in \emph{Neural
  Networks, 1988., IEEE International Conference on},\hskip 1em plus 0.5em
  minus 0.4em\relax IEEE, 1988,  325--333. [Online]. Available:
  \url{http://dx.doi.org/10.1109/ICNN.1988.23864}
\BIBentrySTDinterwordspacing

\bibitem{bishop_neural_add_noise_1995}
\BIBentryALTinterwordspacing
C.~M. Bishop, ``Training with {Noise} is {Equivalent} to {Tikhonov}
  {Regularization},'' \emph{Neural Computation} {\bf 7}, no.~1,  108--116,
  (Jan. 1995). [Online]. Available:
  \url{http://dx.doi.org/10.1162/neco.1995.7.1.108}
\BIBentrySTDinterwordspacing

\bibitem{neural_an_add_noise_1996}
\BIBentryALTinterwordspacing
G.~An, ``\BIBforeignlanguage{en}{The {Effects} of {Adding} {Noise} {During}
  {Backpropagation} {Training} on a {Generalization} {Performance}},''
  \emph{\BIBforeignlanguage{en}{Neural Computation}} {\bf 8}, no.~3,  643--674,
  (Apr. 1996). [Online]. Available:
  \url{http://dx.doi.org/10.1162/neco.1996.8.3.643}
\BIBentrySTDinterwordspacing

\bibitem{zur_neural_add_noise_2009}
\BIBentryALTinterwordspacing
R.~M. Zur, Y.~Jiang, L.~L. Pesce, and K.~Drukker,
  ``\BIBforeignlanguage{en}{Noise injection for training artificial neural
  networks: {A} comparison with weight decay and early stopping},''
  \emph{\BIBforeignlanguage{en}{Med. Phys.}} {\bf 36}, no.~10,  4810, (2009).
  [Online]. Available: \url{http://dx.doi.org/10.1118/1.3213517}
\BIBentrySTDinterwordspacing

\bibitem{AlvarezSNRwithPileup2014}
\BIBentryALTinterwordspacing
R.~E. Alvarez, ``Signal to noise ratio of energy selective x-ray photon
  counting systems with pileup,'' \emph{Med. Phys.} {\bf 41}, no.~11,  111909,
  (2014). [Online]. Available: \url{http://dx.doi.org/10.1118/1.4898102}
\BIBentrySTDinterwordspacing

\bibitem{Overdick2008}
\BIBentryALTinterwordspacing
M.~Overdick, C.~Baumer, K.~J. Engel, J.~Fink, C.~Herrmann, H.~Kruger, M.~Simon,
  R.~Steadman, and G.~Zeitler, ``{Towards direct conversion detectors for
  medical imaging with X-rays},'' \emph{IEEE Trans. Nucl. Sci.} {\bf NSS08},
  1527--1535, (2008). [Online]. Available:
  \url{http://dx.doi.org/10.1109/NSSMIC.2008.4775117}
\BIBentrySTDinterwordspacing

\bibitem{Barrett2003Ch11}
H.~H. Barrett and K.~Myers, \emph{{Foundations of Image Science}},\hskip 1em
  plus 0.5em minus 0.4em\relax Hoboken, NJ: Wiley-Interscience, 2003, Ch.11.

\bibitem{ZimmermanCompareAtable2iterPMB2014}
\BIBentryALTinterwordspacing
K.~C. Zimmerman, E.~Y. Sidky, and T.~Gilat~Schmidt, ``Experimental study of two
  material decomposition methods using multi-bin photon counting detectors,''
  \emph{Proc. SPIE} {\bf 9033},  90\,333G--6, (2014). [Online]. Available:
  \url{http://dx.doi.org/10.1117/12.2043679}
\BIBentrySTDinterwordspacing

\bibitem{touch_neural_SPIEMedimg_2016}
\BIBentryALTinterwordspacing
M.~Touch, D.~P. Clark, W.~Barber, and C.~T. Badea, ``Novel approaches to
  address spectral distortions in photon counting x-ray {CT} using artificial
  neural networks,'' in \emph{{Proceedings SPIE}}, D.~Kontos, T.~G. Flohr, and
  J.~Y. Lo, Eds., Apr. 2016,  97835P. [Online]. Available:
  \url{http://proceedings.spiedigitallibrary.org/proceeding.aspx?doi=10.1117/12.2217037}
\BIBentrySTDinterwordspacing

\bibitem{alvarez2013dimensionality}
\BIBentryALTinterwordspacing
R.~E. Alvarez, ``Dimensionality and noise in energy selective x-ray imaging,''
  \emph{Med. Phys.} {\bf 40}, no.~11,  111909, (2013). [Online]. Available:
  \url{http://dx.doi.org/10.1118/1.4824057}
\BIBentrySTDinterwordspacing

\bibitem{Alvarez1979}
\BIBentryALTinterwordspacing
R.~E. Alvarez and E.~J. Seppi, ``{A comparison of noise and dose in
  conventional and energy selective computed tomography},'' \emph{{IEEE} Trans.
  Nucl. Sci.} {\bf NS-26},  2853--2856, (1979). [Online]. Available:
  \url{http://dx.doi.org/10.1109/TNS.1979.4330549}
\BIBentrySTDinterwordspacing

\bibitem{qrm-ctiodine}
\BIBentryALTinterwordspacing
{QRM-Gmbh}, ``{QRM-CTiodine}.'' [Online]. Available:
  \url{http://www.qrm.de/content/pdf/QRM-CTiodine.pdf}
\BIBentrySTDinterwordspacing

\bibitem{taguchi_modeling_2011}
\BIBentryALTinterwordspacing
K.~Taguchi, M.~Zhang, E.~C. Frey, X.~Wang, J.~S. Iwanczyk, E.~Nygard, N.~E.
  Hartsough, B.~M.~W. Tsui, and W.~C. Barber, ``Modeling the performance of a
  photon counting x-ray detector for {CT:} energy response and pulse pileup
  effects,'' \emph{Med. Phys.} {\bf 38},  1089--1102, (2011). [Online].
  Available: \url{http://dx.doi.org/10.1118/1.3539602}
\BIBentrySTDinterwordspacing

\bibitem{yu_fessler_PMB_2000}
\BIBentryALTinterwordspacing
D.~F. Yu and J.~A. Fessler, ``Mean and variance of single photon counting with
  deadtime,'' \emph{Phys. Med. Biol.} {\bf 45},  2043--2056, (2000). [Online].
  Available: \url{http://dx.doi.org/10.1088/0031-9155/45/7/324}
\BIBentrySTDinterwordspacing

\bibitem{wang_pulse_2011}
\BIBentryALTinterwordspacing
A.~S. Wang, D.~Harrison, V.~Lobastov, and J.~E. Tkaczyk, ``Pulse pileup
  statistics for energy discriminating photon counting x-ray detectors,''
  \emph{Med. Phys.} {\bf 38},  4265--4275, (2011). [Online]. Available:
  \url{http://dx.doi.org/10.1118/1.3592932}
\BIBentrySTDinterwordspacing

\bibitem{PapoulisAthanasios1965}
{Papoulis, Athanasios}, \emph{{Probability, Random Variables, and Stochastic
  Processes}},\hskip 1em plus 0.5em minus 0.4em\relax McGraw-Hill, 1965.

\bibitem{KayV1Chapter3}
S.~M. Kay, \emph{{Fundamentals of Statistical Signal Processing, Volume I:
  Estimation Theory }},\hskip 1em plus 0.5em minus 0.4em\relax Upper Saddle
  River, NJ: Prentice Hall PTR, 1993, Ch. 3.

\bibitem{KayV1Sec4_5}
S.~M. Kay, \emph{{Fundamentals of Statistical Signal Processing, Volume I:
  Estimation Theory }},\hskip 1em plus 0.5em minus 0.4em\relax Upper Saddle
  River, NJ: Prentice Hall PTR, 1993, Sec. 4.5.

\bibitem{Boone1997}
\BIBentryALTinterwordspacing
J.~M. Boone and J.~A. Seibert, ``{An accurate method for computer-generating
  tungsten anode x-ray spectra from 30 to 140 kV},'' \emph{Med. Phys.} {\bf
  24},  1661--70, (1997). [Online]. Available:
  \url{http://dx.doi.org/10.1118/1.597953}
\BIBentrySTDinterwordspacing

\bibitem{Hubbell1999}
\BIBentryALTinterwordspacing
J.~H. Hubbell, ``{Review of photon interaction cross section data in the
  medical and biological context},'' \emph{Phys. Med. Biol.} {\bf 44},
  R1--R22, (1999). [Online]. Available:
  \url{http://dx.doi.org/10.1088/0031-9155/44/1/001}
\BIBentrySTDinterwordspacing

\bibitem{liu2016char-si-det-photon-count2016}
\BIBentryALTinterwordspacing
X.~Liu, ``Characterization and energy calibration of a silicon-strip detector
  for photon-counting spectral computed tomography,'' Ph.D. dissertation, KTH
  Royal Institute of Technology, 2016. [Online]. Available:
  \url{http://www.diva-portal.org/smash/record.jsf?pid=diva2:967342}
\BIBentrySTDinterwordspacing

\end{thebibliography}
\end{document}